\documentclass{article}

\usepackage{epsfig}

\newtheorem{thm}{Theorem}
\newtheorem{lem}{Lemma}
\newtheorem{tab}{Table}
\newtheorem{fig}{Figure}

\newtheorem{automaton}{Automaton}

\def\leurre{\noindent\leftskip0pt\small\baselineskip 10pt}

\def\encadre#1#2{%
\setbox100=\hbox{\kern#1{#2}\kern#1}
\dimen100=\ht100 \advance \dimen100 by #1
\dimen101=\dp100 \advance \dimen101 by #1
\setbox100=\hbox{\vrule height \dimen100 depth \dimen101\box100\vrule}
\setbox100=\vbox{\hrule\box100\hrule}
\advance \dimen100 by .4pt \ht100=\dimen100
\advance \dimen101 by .4pt \dp100=\dimen101
\box100
\relax
}

\def\ligne#1{\hbox to \hsize{#1}}
\def\PlacerEn#1 #2 #3 {\rlap{\kern#1\raise#2\hbox{#3}}}

\def\grostrait{\ligne{\vrule height 1pt depth 1pt width \hsize}}
\def\demitrait{\ligne{\vrule height 0.5pt depth 0.5pt width \hsize}}

\def\reunion{\mathop{\cup}}

\begin{document}

\title{Iterative pushdown automata and hyperbolic contour words}

\author{Maurice Margenstern} 

\maketitle

\begin{center}\small
Laboratoire d'Informatique Th\'eorique et Appliqu\'ee, EA 3097,\\
Universit\'e de Metz, I.U.T. de Metz,\\
D\'epartement d'Informatique,\\
\^Ile du Saulcy,\\
57045 Metz Cedex, France,\\
{\it email}{margens@univ-metz.fr}
\end{center}


\vskip 10pt
\begin{abstract}
In this paper, we give an application of iterated pushdown automata to contour words of
balls and two other domains in infinitely many tilings of the hyperbolic plane. We also 
give a similar application
for the tiling $\{5,3,4\}$ of the hyperbolic $3D$ space and for the tiling $\{5,3,3,4\}$
of the hyperbolic $4D$ space as well.
\end{abstract}

{\bf keywords}:
pushdown automata, iterated pushdown automata, tilings, hyperbolic plane, 
tessellations

ACM-class: F.2.2., F.4.1, I.3.5

\vskip 10pt

\def\cqfd{\hbox{\kern 2pt\vrule height 6pt depth 2pt width 8pt\kern 1pt}}
\def\Hii{$I\!\!H^2$}
\def\Hiii{$I\!\!H^3$}
\def\Hiv{$I\!\!H^4$}
\def\norm{\hbox{$\vert\vert$}}
\section{Introduction}

   Iterated pushdown automata were introduced in~\cite{greibach,maslov} and we refer
the reader to~\cite{fratani-senizergues} for references and for the connection of this
topic with sequences of rational numbers. By their definition, iterated pushdown 
automata are more powerful than standard pushdown automata but they are far less powerful
than Turing machines. As Turing machines can be similated by a finite automaton with
two independent stacks, iterated pushdown automata can be viewed as an intermediate
device.

   In this paper, we show an application of this device to the characterization
of contour words of a family of bounded domains in many tilings of the hyperbolic
plane. We do the same kind of application for a tiling of the hyperbolic $3D$ space
and for another one in the hyperbolic $4D$ space. These two latter applications cannot
be generalized to any dimension as, starting from dimension~5, there is no tiling of the
hyperbolic space which would be a tessellation generated by a regular polytope.

   In section~2, we remember the definition of iterated pushdown automata with
an application to the computation of the recognition of words of the form $a^{f_n}$, where
$\{f_n\}_{n\in I\!\!N}$ is the Fibonacci sequence with $f_0=f_1=1$. This sequence
will always denoted by $\{f_n\}_{n\in I\!\!N}$ in all the paper.

   In section~3, we remind the reader several features and properties on tilings of
the hyperbolic plane.

   In section~4, we define the contour words which we are interested in and we
construct iterated pushdown automata which recognize them for the case
of the pentagrid and the heptagrid, {\it i.e.} the tilings $\{5,4\}$ and $\{7,3\}$ 
of the hyperbolic plane. We also extend these results to infinitely many tilings of 
the hyperbolic plane.

   In section~5, we extend the result to two tilings of the $3D$ and~$4D$ hyperbolic spaces.

\section{Iterated pushdown automata}

   In this section, we fix the notations which will be used in the paper. We follow
the notations of \cite{fratani-senizergues}.

\subsection{Iterated pushdown stores}

   This data structure is defined by induction, as follows:

   {\leftskip 20pt\parindent 0pt
   0-pds($\Gamma$) = $\{\epsilon\}$\vskip 2pt
   $k$+1-pds($\Gamma$) = $(\Gamma[k$-pds$(\Gamma)])^*$\vskip 2pt
   it-psd($\Gamma$) = $\reunion_{k}k$-pds$(\Gamma)$
   \par}

   The elements of a $k$+1-pds($\Gamma$) structure are $k$-pds$(\Gamma)$ structures
and each element is labelled by a letter of~$\Gamma$. A $k$-pds$(\Gamma)$ structure
will often be called a $k$-level store, for short. When $k$ is fixed, we speak of
{\it outer} stores and of {\it inner} stores in a relative way: an $i$-level store
is {\bf outer} than a $j$-one if and only if $i< j$. In the same situation, the
$j$-level store is {\bf inner} than the $i$-one.

   We define functions and operations on $k$-level stores, by induction 
on~$k$.

   From the above definition, we get that a $k$+1-level store~$\omega$ can be uniquely 
represented in the form:

\ligne{\hfill $\omega = A[flag].rest$,\hfill}

\noindent
where $A\in\Gamma$, $flag$ is a $k$-level store and $rest$ is $k$+1-store. Moreover,
if $\ell$ is the number of elements of~$rest$, the number of elements of~$\omega$
is $\ell$+1.

   A first operation consists in defining the generalization of the standard notion of 
{\it top symbol} in an ordinary pushdown structure. This is performed by the function
$topsym$ defined by:

   {\leftskip 20pt\parindent 0pt
   $topsym(\epsilon) = \epsilon$\vskip 2pt
   $topsym(A[flag].rest) = A.topsym(flag)$
   \par}

It is important to remark that $topsym$ is the single direct access to all inner stores
of a $k$-level store. In other words, for any inner store, only its topmost symbol can
be accessed and when this inner store is in the top of the outest store. 

   Also note that the $topsym$ function performs a {\bf reading}. There are two 
families of {\bf writing} operations, also concerning the elements visible from 
the $topmost$ function only.

   The first one consists of the $pop$ operations defined by the following induction:

   {\leftskip 20pt\parindent 0pt
   $pop_j(\epsilon)$ is undefined \vskip 2pt
   $pop_{j+1}(A[flag].rest) = A.[pop_j(flag)].rest$
   \par}

   The second family consists of the $push$ operations defined by the following induction:

   {\leftskip 20pt\parindent 0pt
   $push_1(\gamma)(\epsilon) = \gamma$, for $\gamma\in\Gamma$ \vskip 2pt
   $push_j(\gamma)(\epsilon)$ is undefined for $j> 1$ \vskip 2pt
   $push_{j+1}(w)(A[flag].rest) = w_1[flag]..w_k[flag].rest$,
where $w = w_1..w_k$, with $w_i\in \Gamma$ for $1\le i\le k$
   \par}

\subsection{Iterated pushdown automata}

   Intuitevely, the definition is very close to the traditional one of standard
non-deterministic standard automata. A $k$-iterated pushdown automaton is defined
by giving the following data:

{\leftskip 20pt\parindent 0pt
- a finite set of states, $Q$;\vskip 2pt
- an input finite alphabet $\Sigma$;\vskip 2pt
- a store finite alphabet $\Gamma$;\vskip 2pt
- a transition function $\delta$ from $Q\times\Sigma\cup\{\epsilon\}\times\Gamma^k$ into
a finite set of instructions of the form $(q,${\bf op}$)$, where $q$~is a state
and {\bf op} is a pop- or a push-operation as described in the previous sub-section. 
\par}
  
   We also assume that there is an initial state denoted by~$q_0$ and that the initial
state of the store is $Z[\epsilon]$, where $Z$~is a fixed in advance symbol of~$\Gamma$.
Note that we allow $\epsilon$-transition which play a key role.

   A configuration is a word of the form $(q,w,\omega)$, where $q$~is the current state 
of the automaton, $w$ is the current word and $\omega$ is the current $k$-level store of the 
automaton. A computational step of the automaton allows to go from one configuration
to another by the application of one transition. In order to be applied, $q$ state of the 
automaton must be that of the transition, the first letter of~$w$ must be the symbol
of~$\Sigma$ in the transition if any, and $topsym(\omega)$ must be the word of
$\Gamma^k$ in the transition if any. A word $w$ is accepted if and only there
is a sequence of computational steps starting from $(q_0,w,Z[\epsilon])$ to
a first configuration of the form $(q,\epsilon,\epsilon)$. The language recognized
by a $k$-iterated pushdown automaton is the set of words in $\Sigma^*$ which are
accepted by the automaton.

   As an illustrative example of the working of such an automaton, we shall take the 
2-pushdown automaton given in~\cite{fratani-senizergues}, which recognizes the words of 
the form $a^{f_n}$ where $\{f_n\}_{n\in I\!\!N}$ is the Fibonacci sequence, see
the figure of automaton~\ref{autom1}.

   Now, it is not difficult to prove the following lemma:

\begin{lem}\label{lem1}
We have the following relations, for any nonnegative~$k$:
\vskip 5pt
   $(q_0,a^{f_k},X_2[F^k].\omega) \Rightarrow_\delta^* (q_0,\epsilon,\omega)$\vskip 2pt
   $(q_0,a^{f_{k+1}},X_1[F^k].\omega) \Rightarrow_\delta^* (q_0,\epsilon,\omega)$
\end{lem}

\noindent
Proof. It is performed by induction whose basic case $k=0$ is easy.
If we start from $(q_0,a^{f_{k+1}},X_1[F^k].\omega)$, we have the following
derivation:

{\leftskip 20pt\parindent 0pt
 $(q_0,a^{f_{k+2}},X_1[F^{k+1}].\omega) \vdash  (q_1,a^{f_{k+2}},X_1[F^k].\omega)$\vskip 2pt
 $\vdash  (q_0,a^{f_{k+2}},X_1[F^k].X_2[F^k].\omega)
 \vdash (q_0,a^{f_k},X_2[F^k].\omega)$
\par}

\noindent
by induction hypothesis as $f_{k+2} = f_{k+1}+f_k$. And, again by induction hypothesis:

{\leftskip 20pt\parindent 0pt
 $(q_0,a^{f_k},X_2[F^k].\omega) \vdash (q_0\epsilon,\omega)$
\par}

   Similarly,

{\leftskip 20pt\parindent 0pt
 $(q_0,a^{f_{k+1}},X_2[F^{k+1}].\omega) \vdash  (q_2,a^{f_{k+1}},X_2[F^k].\omega)
 \vdash (q_0,a^{f_{k+1}},X_1[F^k].\omega)$\vskip 2pt 
 $\vdash  (q_0,\epsilon,\omega)$,
\par}

\noindent
by induction hypothesis.

   Let $a^m$ be the initial word. With the first two transitions, we guess an integer~$k$
such that $m = f_k$ if any. Then we arrive to the configuration $(q_0,a^m,Z[F^k])$.
Next, we have:

{\leftskip 20pt\parindent 0pt
$(q_0,a^m,Z[F^k]) \vdash (q_0,a^m,X_2[F^k])$.
\par}

\noindent
And by the lemma, we proved that $(q_0,a^m,X_2[F^k])\vdash (q_0,\epsilon,\epsilon)$ and
so, the word is accepted.

   We can see that if $m=f_k$ and if we guessed a wrong~$k$, then either the word is
not empty when the store vanishes, and we cannot restore it, or the word is empty as the
store is not. This also shows that if $m\not= f_k$, as there is in this case a unique
$k$ such that $f_k<m<f_{k+1}$, we always have either an empty word an a non-empty store
or an empty store with a nonempty word, whatever the guess.\cqfd

\vtop{
\begin{automaton}\label{autom1}
\leurre
The $2$-pushdown automaton recognizing the Fibonacci sequence.
\end{automaton}
\vspace{-12pt}
\grostrait
\vskip 0pt
three states: $q_0$, $q_1$ and~$q_2$; input word in $\{a\}^*$; 
$\Gamma = \{Z,X_1,X_2,F\}$;
\vskip 0pt\noindent
initial state: $q_0$; initial stack: $Z[\epsilon]$; transition function $\delta$: 
\vskip 0pt
\vspace{-4pt}
\demitrait
\vskip 0pt
{\leftskip 20pt\parindent 0pt
$\delta(q_0,\epsilon,Z) = \{(q_0,push_2(F)),(q_0,push_1(X_2))\}$\vskip 2pt
$\delta(q_0,\epsilon,ZF) = \{(q_0,push_2(FF)),(q_0,push_1(X_2))\}$\vskip 2pt
$\delta(q_0,\epsilon,X_1F) = (q_1,pop_2)$\vskip 2pt
$\delta(q_0,\epsilon,X_2F) = (q_2,pop_2)$\vskip 2pt
$\delta(q_0,a,X_1) = (q_0,pop_1)$\vskip 2pt
$\delta(q_0,a,X_2) = (q_0,pop_1)$\vskip 2pt
$\delta(q_1,\epsilon,X_1F) = (q_0,push_1(X_1X_2))$\vskip 2pt
$\delta(q_2,\epsilon,X_2F) = (q_0,push_1(X_1))$\vskip 2pt
$\delta(q_1,\epsilon,X_1) = (q_0,push_1(X_1X_2))$\vskip 2pt
$\delta(q_2,\epsilon,X_2) = (q_0,push_1(X_1))$\vskip 2pt
\par}
\vskip 0pt
\demitrait
}
\vskip 5pt

\section{The tilings of the hyperbolic plane}

   We assume that the reader is a bit familiar with hyperbolic geometry, at least
with its most popular models, the Poincar\'es's half-plane and disc.

   We remember the reader that in the hyperbolic plane, thanks to a well known theorem
of Poincar\'e, there are infinitely many tilings which are generated by tessellation
starting from a regular polygon. This means that, starting from the polygon,
we recursively copy it by reflections in its sides and of the images in their sides.
This family of tilings is defined by two parameters: $p$, the number of sides of
the polygon and $q$, the number of polygons which can be put around a vertex without
overlapping and covering any small enough neighbourhood of the vertex.

\ligne{\hfill}
\setbox110=\hbox{\epsfig{file=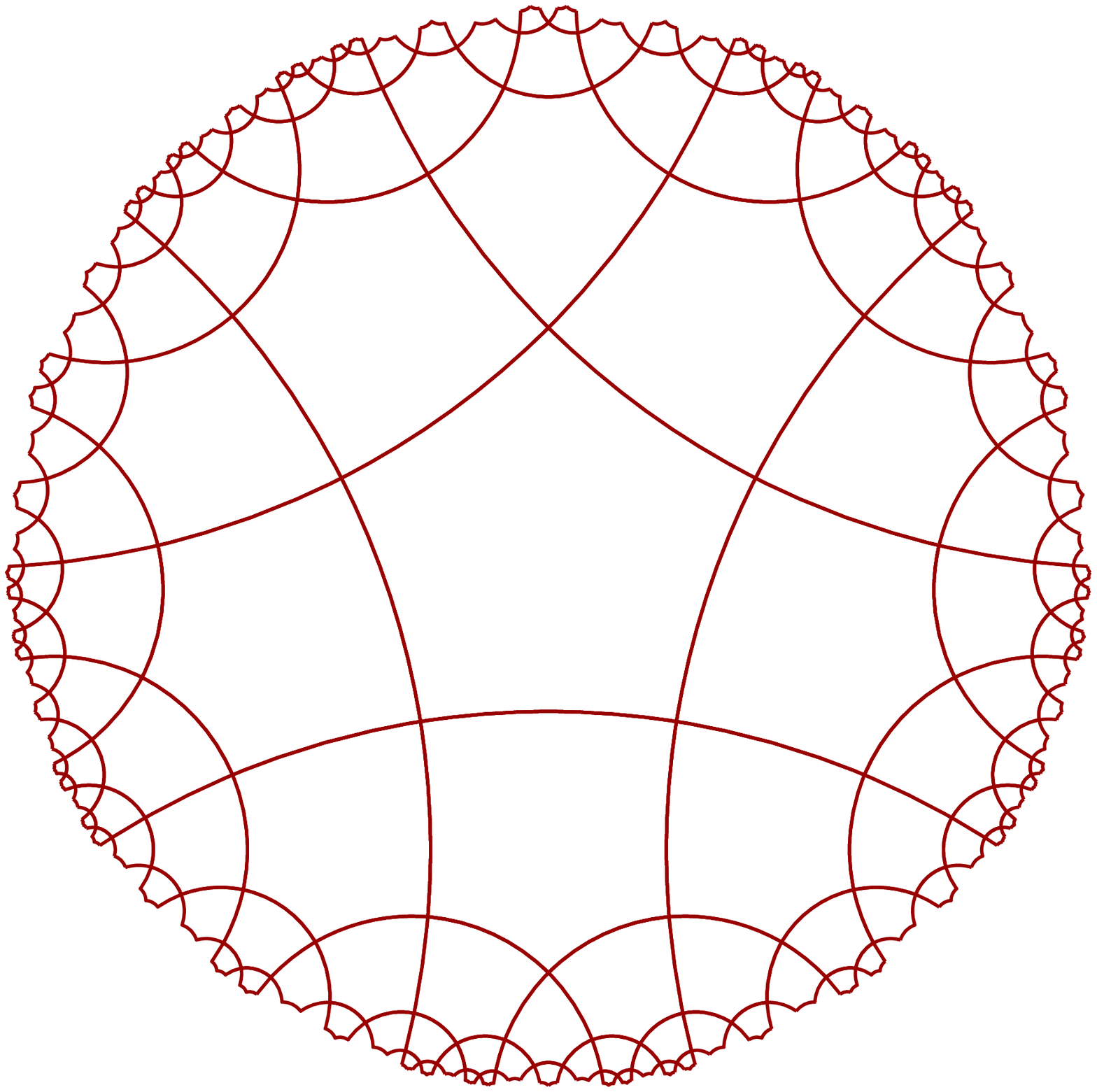,width=160pt}}
\setbox112=\hbox{\epsfig{file=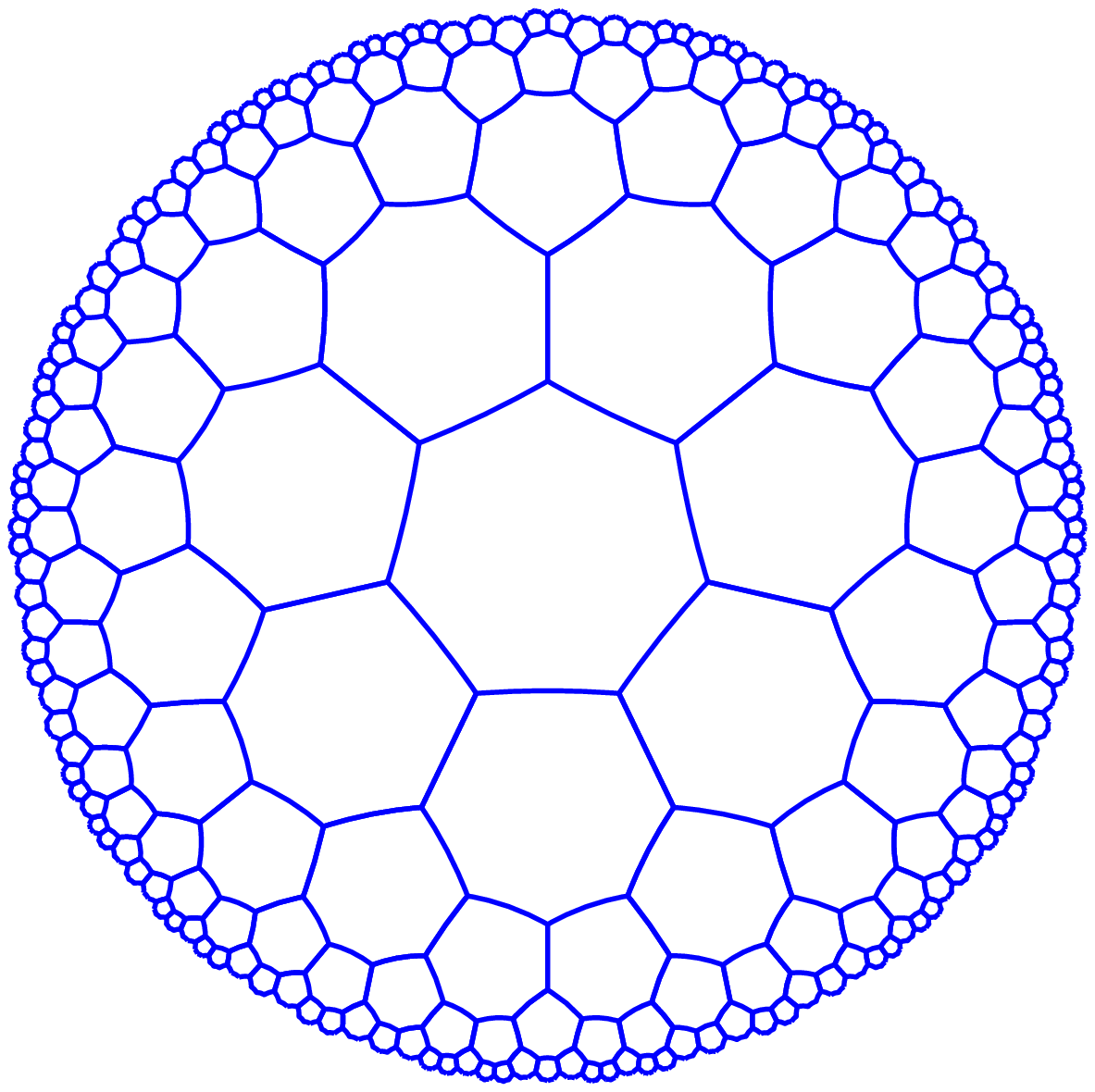,width=160pt}}
\vtop{
\ligne{\hfill
\PlacerEn {-165pt} {0pt} \box110
\PlacerEn {-5pt} {0pt} \box112
\hfill
}
\begin{fig}\label{tilings}
\leurre
Left-hand side: the pentagrid. Right-hand side: the heptagrid. 
\end{fig}
}

\ligne{\hfill}
\setbox110=\hbox{\epsfig{file=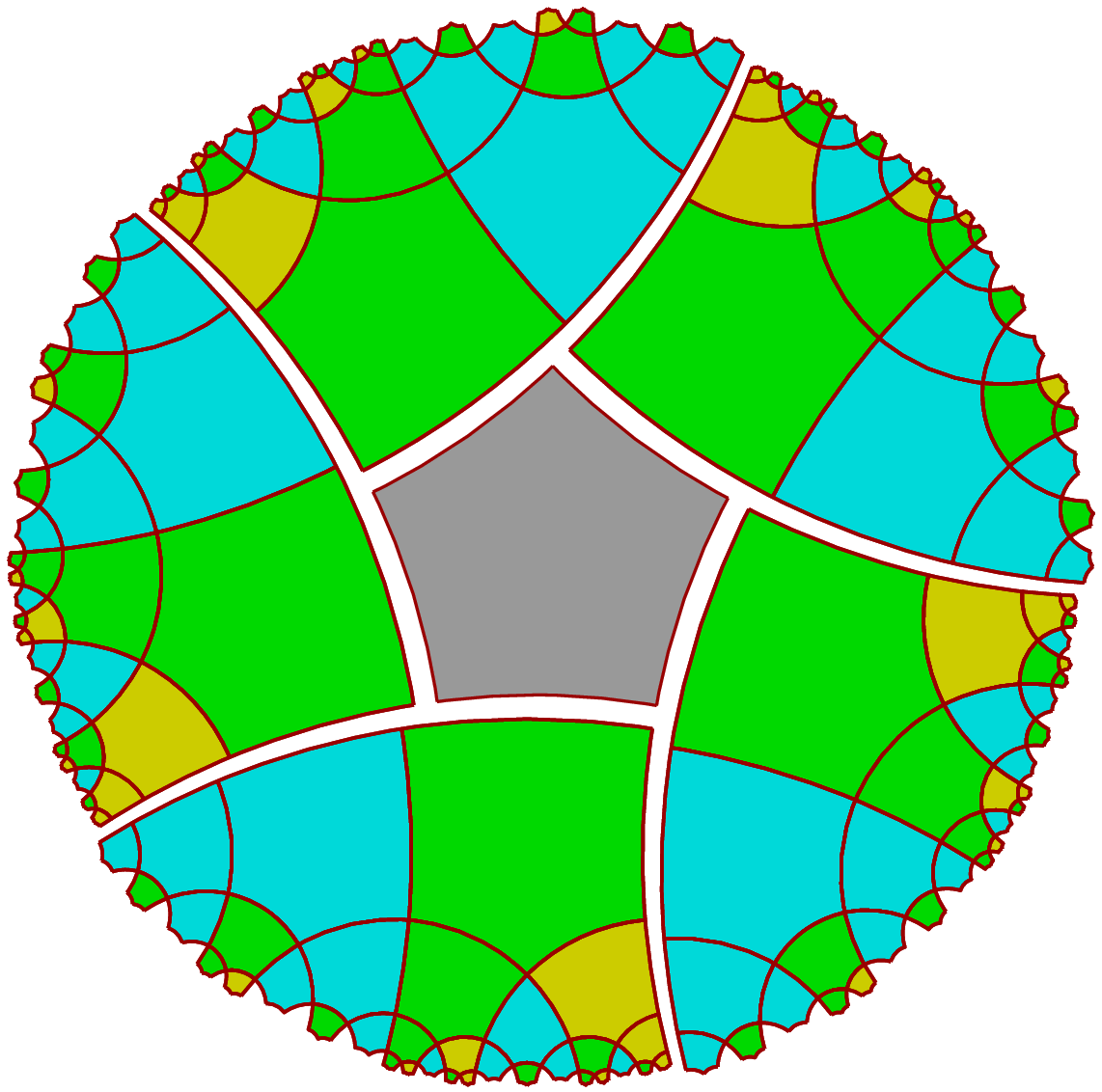,width=155pt}}
\setbox112=\hbox{\epsfig{file=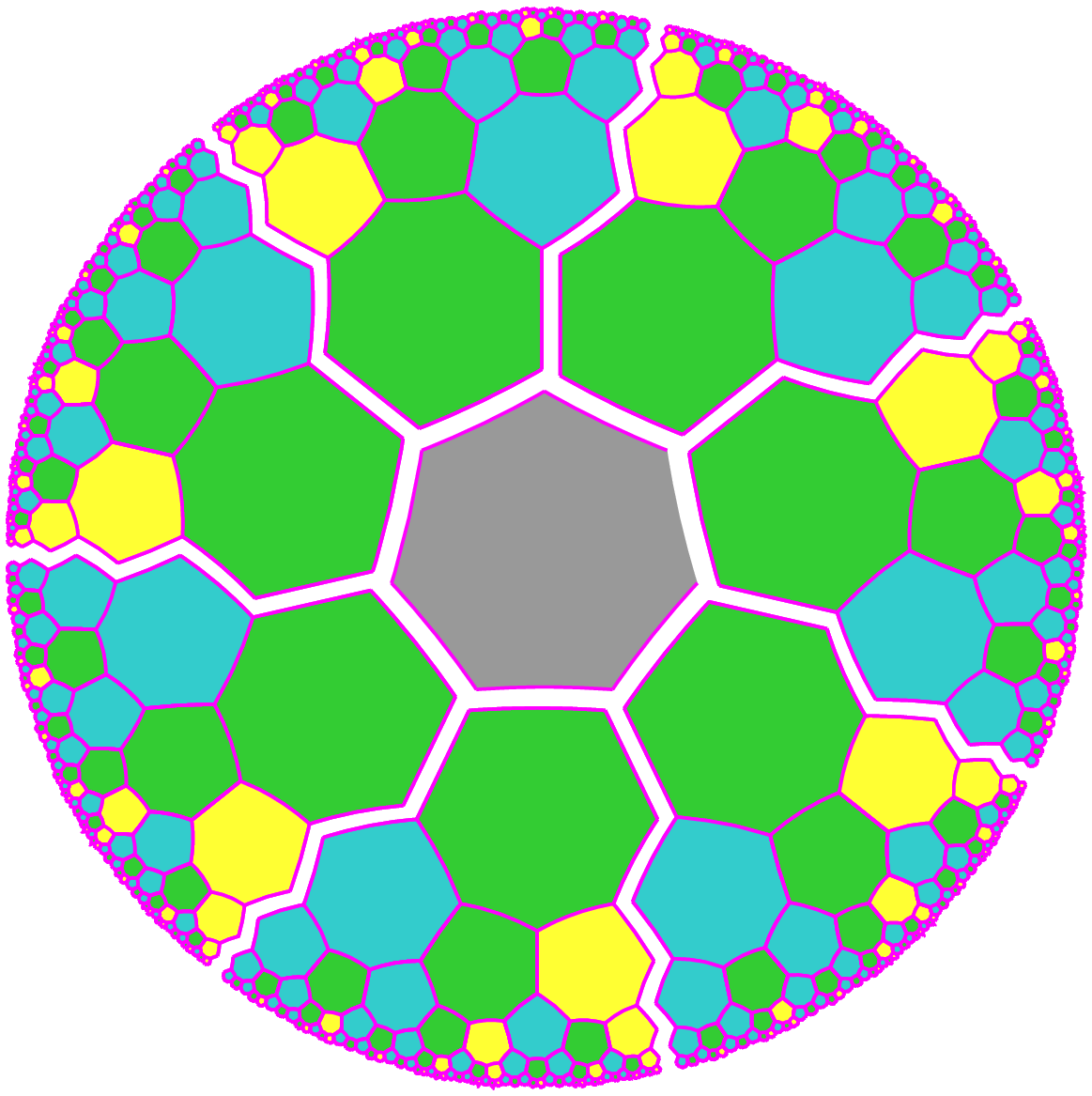,width=155pt}}
\vtop{
\ligne{\hfill
\PlacerEn {-155pt} {0pt} \box110
\PlacerEn {5pt} {0pt} \box112
\hfill
}
\begin{fig}\label{splittings}
\leurre
Left-hand side: the pentagrid. Right-hand side: the heptagrid. 
Note that in both cases, the sectors are spanned by the same tree.
\end{fig}
}

   In order to represent the tilings which we shall consider and the regions whose
contour word will be under study, we shall make use of the Poincar\'e's disc model.
Our illustrations will take place in the {\bf pentagrid} and the {\bf heptagrid},
{\it i.e.} the tilings $\{5,4\}$ and $\{7,3\}$ respectively of the hyperbolic plane.
Below, Figures~\ref{tilings} and~\ref{splittings} will illustrate these tilings.

   From Figure~\ref{tilings},the pentagrid and the heptagrid seem rather different.
However, there is a tight connection between these tilings which can be seen from
Figure~\ref{splittings}. In both pictures of the latter figure, we represent
the tiling by selecting a central tile and then, by displaying as many sectors as
the number of sides of the central tile. In each case, these sectors do not overlap 
and their union together with the central cell gives the tiling of the whole hyperbolic
plane. Now, there is a deeper common point: in both cases, each sector is spanned by
a tree which we call a Fibonacci tree for a reason which will soon be explained.
 
   As proved in \cite{mmJUCSii,mmbook1}, the corresponding tree can be defined
as follows. We distinguish two kinds of nodes, say black nodes, labelled by~$B$, and
white nodes, labelled by~$W$. Now, we get the sons of a node by the following rules:
$B\rightarrow BW$ and $W\rightarrow BWW$, the root of the tree being a white node. 
It is not difficul to see that if the root is on level~0 of the tree, the number
of nodes on the level~$k$ of the tree is $f_{2k+1}$, where $\{f_k\}_{k\in I\!\!N}$
is the Fibonacci sequence with $f_0=f_1=1$.

\vskip-5pt
\ligne{\hfill}
\setbox110=\hbox{\epsfig{file=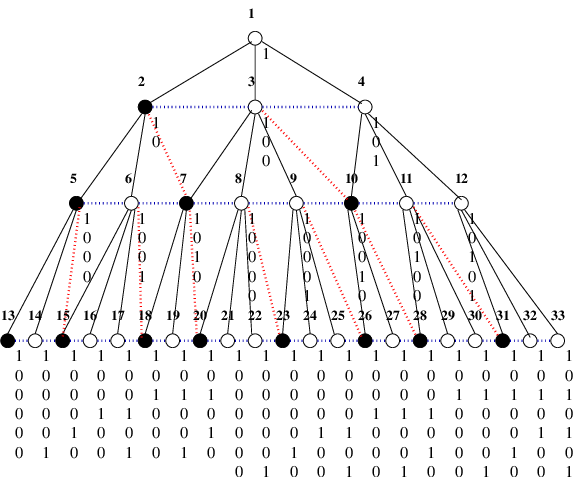,width=250pt}}
\vtop{
\ligne{\hfill
\PlacerEn {-135pt} {10pt} \box110
\hfill
}
\begin{fig}\label{fibotree}
\leurre
The standard Fibonacci tree. The nodes are numbered from the root, from left to right on
each level and level after level. For each node, the figure displays the representation
of the number of the node with respect to the Fibonacci sequence, the representation
avoiding consecutive $1$'s.
\end{fig}
}

   The Fibonacci tree has a lot of nice properties which we cannot discuss here.
In particular, there is a way to locate the tiles of the pentagrid or the heptagrid
very easily thanks to coordinates devised from the properties of the Fibonacci tree,
see \cite{mmJUCSii,mmbook1,mmbook2}.

\section{The contour words}

   Now, we have the tools to define the regions from which we define contour words, and
we also have the tools to define the iterated pushdown automata which recognize them.
   
\subsection{Balls}

    Consider a ball $B_k$ of radius~$k$+1{} in the penta- or the heptagrid. This ball
is the union of the central cell and $\alpha$ truncated sectors, $\alpha=5$ for the pentagrid
and $\alpha=7$ for the heptagrid, each truncated sector being spanned by a Fibonacci
tree up to the level~$k$. Accordingly, the number of tiles which are exactly at the
distance $k$+1 from the central cell is~$\alpha.f_{2k+1}$. Denote the set
of these tiles by~$\partial B_k$.

   The $\alpha$ sectors around the central cell can be numbered from~1 to~$\alpha$
by fixing sector~1 once for all and by counterclockwise turning around the central cell.
We call {\bf contour word} $cw_k$, the word obtained by taking the labels of
the tiles which are on $\partial B_k$, starting from the left-hand side border of sector~1
and by counter clockwise running along $\partial B_k$, until the tile which is on
the right-hand side border of the sector~$\alpha$. It is easy to remark that the
contour work $cw_k$ can be written as $(sw_k)^\alpha$, where $sw_k$ is obtained
by taking the word which is on~$\partial B_k$ from the left-hand side of the sector
to its right-hand side.

\vskip-5pt
\ligne{\hfill}
\setbox110=\hbox{\epsfig{file=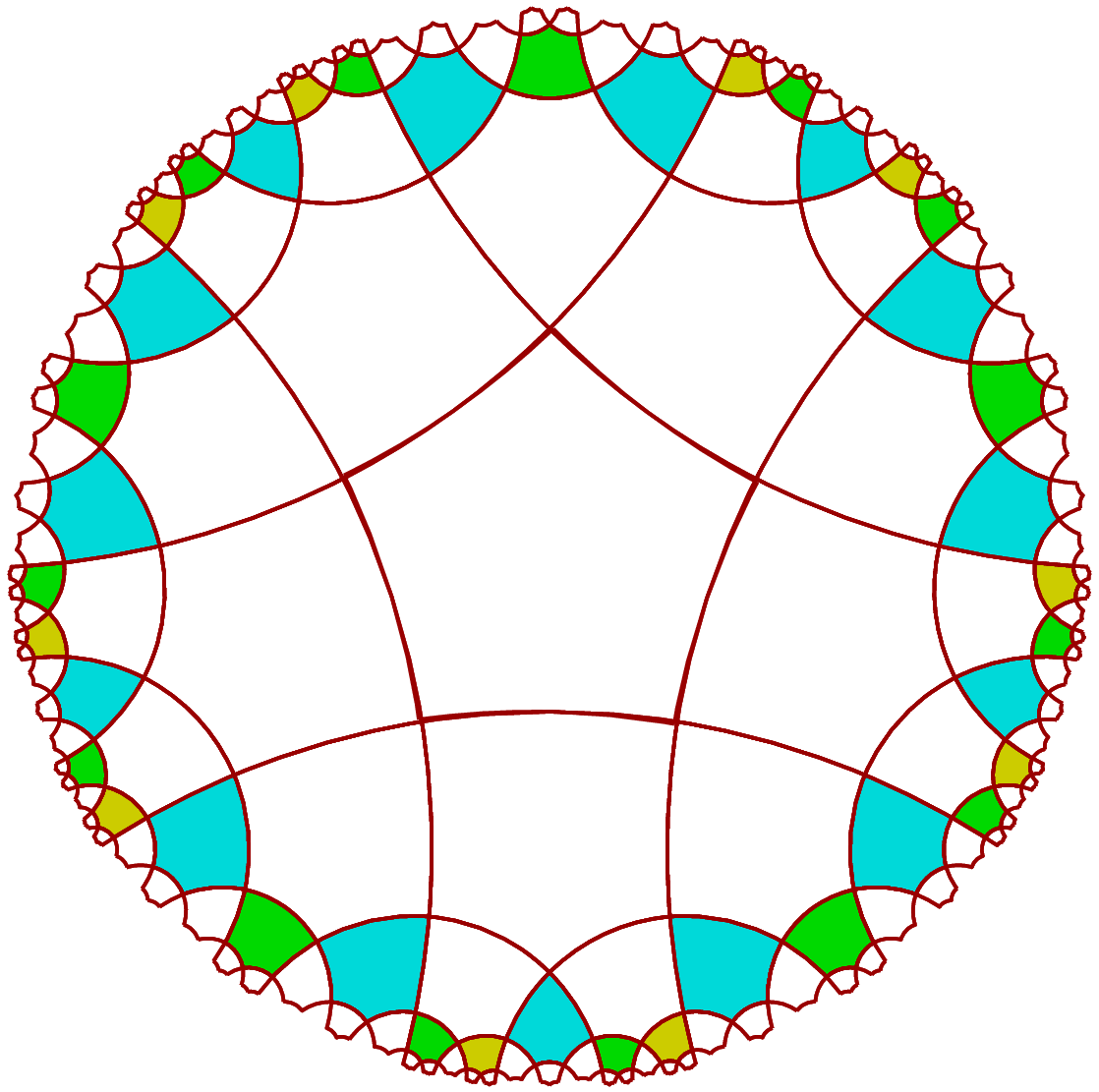,width=155pt}}
\setbox112=\hbox{\epsfig{file=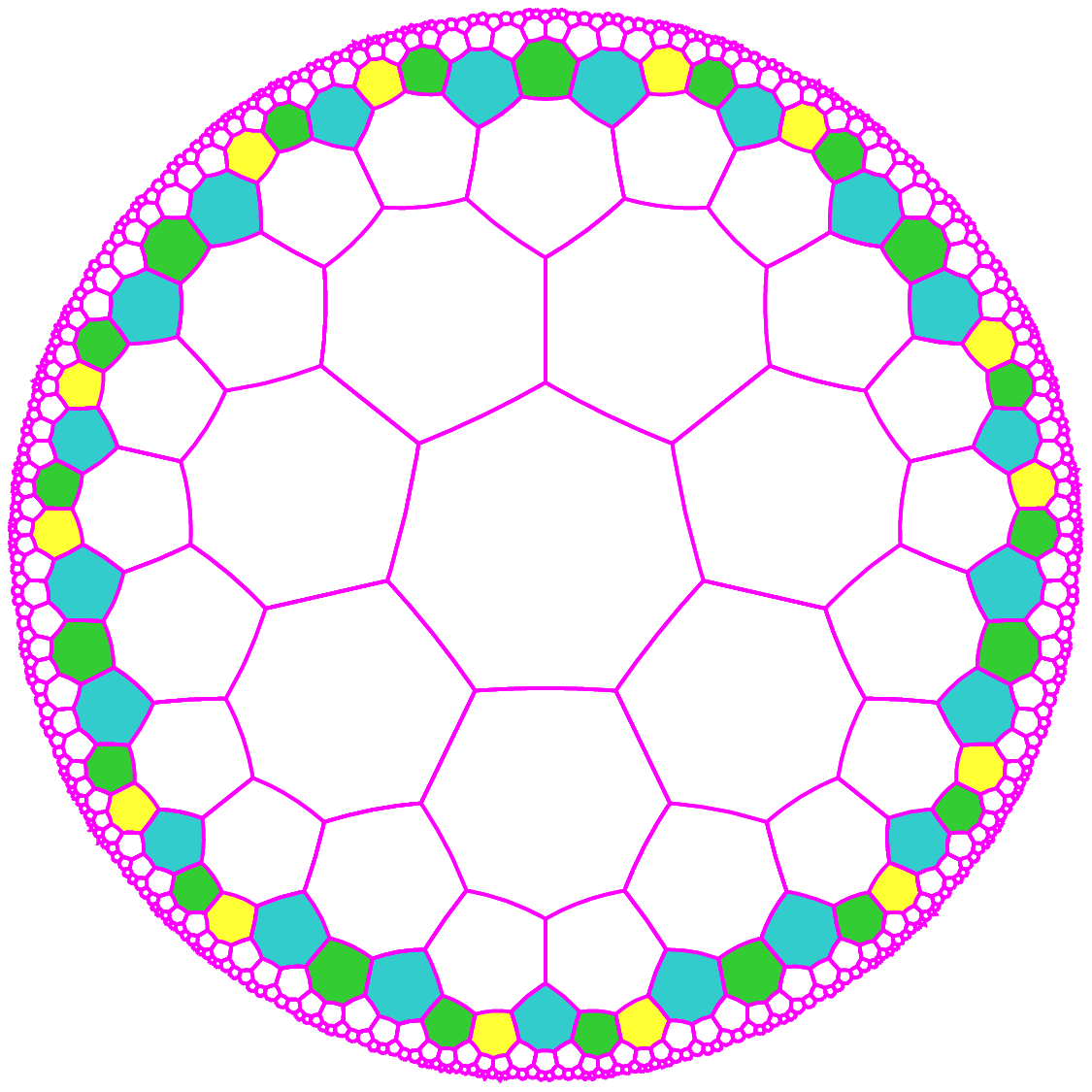,width=155pt}}
\vtop{
\ligne{\hfill
\PlacerEn {-155pt} {0pt} \box110
\PlacerEn {5pt} {0pt} \box112
\hfill
}
\begin{fig}\label{contours}
\leurre
The representation of $\partial B_3$.
Left-hand side: the pentagrid. Right-hand side: the heptagrid. In both cases,
$sw_3= bwbwwbww$.
\end{fig}
}

   Already by the length of the contour words, we can see that the set of all
contour words is not algebraic: it is enough to apply Ogden's pumping lemma.

   Now we have:

\begin{thm}\label{auto57}
The contour words of the pentagrid and those of the heptagrid can be recognized by
a $2$-level pushdown automaton.
\end{thm}
     
\noindent
Proof. Indeed, we can transform Automaton~\ref{autom1} in order to do the job.
Here is the automaton:

\vtop{
\begin{automaton}\label{autom2}
\leurre
The $2$-pushdown automaton recognizing the contour word of a ball in the pentagrid or
in the heptagrid.
\end{automaton}
\vspace{-12pt}
\grostrait
\vskip 0pt
two states: $q_0$ and~$q_1$; input word in $\{b,w\}^*$; 
$\Gamma = \{Z,B,W,F\}$;
\vskip 0pt\noindent
initial state: $q_0$; initial stack: $Z[\epsilon]$; transition function $\delta$: 
\vskip 0pt
\vspace{-4pt}
\demitrait
\vskip 0pt
{\leftskip 20pt\parindent 0pt
$\delta(q_0,\epsilon,Z) = \{(q_0,push_2(F)),(q_0,push_1(W^\alpha))\}$\vskip 2pt
$\delta(q_0,\epsilon,ZF) = \{(q_0,push_2(FF)),(q_0,push_1(W^\alpha))\}$\vskip 2pt
$\delta(q_0,\epsilon,WF) = (q_1,pop_2)$\vskip 2pt
$\delta(q_0,\epsilon,BF) = (q_1,pop_2)$\vskip 2pt
$\delta(q_0,b,B) = (q_0,pop_1)$\vskip 2pt
$\delta(q_0,w,W) = (q_0,pop_1)$\vskip 2pt
$\delta(q_1,\epsilon,WF) = (q_0,push_1(BWW))$\vskip 2pt
$\delta(q_1,\epsilon,BF) = (q_0,push_1(BW))$\vskip 2pt
$\delta(q_1,\epsilon,W) = (q_0,push_1(BWW))$\vskip 2pt
$\delta(q_1,\epsilon,B) = (q_0,push_1(BW))$\vskip 2pt
\par}
\vskip 0pt
\demitrait
}
\vskip 10pt
 
\begin{lem}\label{lem2}
We have the following relations, for any nonnegative~$k$:
\vskip 5pt
   $(q_0,bsw_k,B[F^k].\omega) \Rightarrow_\delta^* (q_0,\epsilon,\omega)$\vskip 2pt
   $(q_0,sw_k,W[F^k].\omega) \Rightarrow_\delta^* (q_0,\epsilon,\omega)$
\end{lem}

\noindent
where $bsw_k$ is the word obtained on the level~$k$ of a Fibonacci tree whose
root is a black node, starting from the left-hand side border to the right-hand side
one.

   The easy proof, srictly parallel to that of Lemma~\ref{lem1}, is left ot the reader.
However, we can give an idea of the automaton which will better convince the
reader. 
   
   In fact, the automaton can be seen as a device which traverse the tree in a depth first
way. To this purpose, starting from the root whose height is~$k$, the automaton
puts on the outer store the nodes it takes, always going to left first. For each node,
the automaton puts the labels of the sons of the node together with the height of each
sons which is stored as an inner store. As the height of the sons is reduced by~1 with 
respect to that of the father, this obtained by a simple popping of the inner store.
This explains both the construction of the automaton and its correctness.  

   Of course, the figure of automaton~\ref{autom2} provides us with two automata: one for the
pentagrid and one for the heptagrid.\cqfd

   We remark from the proof that the result can be extended to any tree with a finite
branching where the degree of the nodes can be defined by a fixed set of rules. A similar
automaton can then be easily deduced. We refer the reader to~\cite{fratani-senizergues} 
for more information on the connection between iterated pushdown automata and trees.

   As proved in \cite{mmgsJUCS,mmbook1}, the tilings $\{p,4\}$ and $\{p$+$2,3\}$ of the
hyperbolic plane are spanned by the same tree, which can be see as a generalization of
the Fibonacci tree. The tree has two kinds of nodes, again black and white, labelled
by~$B$ and~$W$ respectively, as for the Fibonacci tree, and now, the rules are:
\vskip 5pt
\ligne{\hfill
$W\rightarrow BW^{p-3}$ and $B\rightarrow BW^{p-4}i.$\hfill$(R)$\hskip 20pt
}

   We can define balls in the tilings $\{p,4\}$ and $\{p$+$2,3\}$ as in the case of
the pentagrid or the heptagrid: it is in fact a general definition. Now, we can define
also the border of a ball and the contour word which is defined as in the case
of the pentagrid and of the heptagrid. 

   Now, considering the transitions of automaton~\ref{autom2}, it is easy to change
them in order to obtain a 2-pushdown automaton which exactly recognizes the contour
words of the balls of the tilings $\{p,4\}$ and $\{p$+$2,3\}$. And so, we can state:

\begin{thm}\label{pp+2}
   There is a $2$-pushdown automaton which recognizes
exactly the contour words of the balls of the tilings $\{p,4\}$ and $\{p$+$2,3\}$.
\end{thm}

\subsection{Sectors}

   Now, we can also consider truncated sectors as a region. We can define two kinds
of truncated sectors which were already studied in~\cite{mmbook1} as {\it quarters}
and {\it bars} in the case of the pentagrid.

   A truncated sector of the first kind, we shall say a {\bf white} $k$-{\bf sector},
consists of the tiles which belong to a Fibonacci tree rooted at a white tile up to 
the level~$k$, this level being included. Similarly, we define a {\bf black} 
$k$-{\bf sector} using a Fibonacci tree rooted at a black node. It is plain that a white 
and a black $k$-sector can be defined indifferently in the pentagrid or in the heptagrid.
Moreover, these notions can be extended to the tilings $\{p,4\}$ and $\{p$+$2,3\}$,
replacing the Fibonacci tree by the tree defined by the rules~$(R)$.  

   The contour word of a $k$-sector is defined by its border, as in the case of a ball.
Let $W_k$ be a white $k$-sector. Then, its border, $\partial W_k$ is defined as the set of
tiles which are on the leftmost branch of the tree spanning the sector, on its rightmost 
branch or on the level~$k$ of the tree, see Figure~\ref{cont_Wsectors}. 

   Similarly, if $B_k$ is black $k$-sector, its border, $\partial B_k$, is defined as the
set of tiles which are on the leftmost branch of the tree spanning the sector, on its
rightmost branch or on the level~$k$ of the tree, see Figure~\ref{cont_Bsectors}. 

   Now that we defined the contour words attached to $\partial W_k$ and $\partial B_k$,
we can prove the following result:

\begin{thm}\label{thm3}
There is a $2$-iterated pushdown automaton which recognizes the contour word of all
$\partial W_k$'s as well as another one to recognize the contour words of all 
$\partial B_k$'s.
\end{thm}

\ligne{\hfill}
\setbox110=\hbox{\epsfig{file=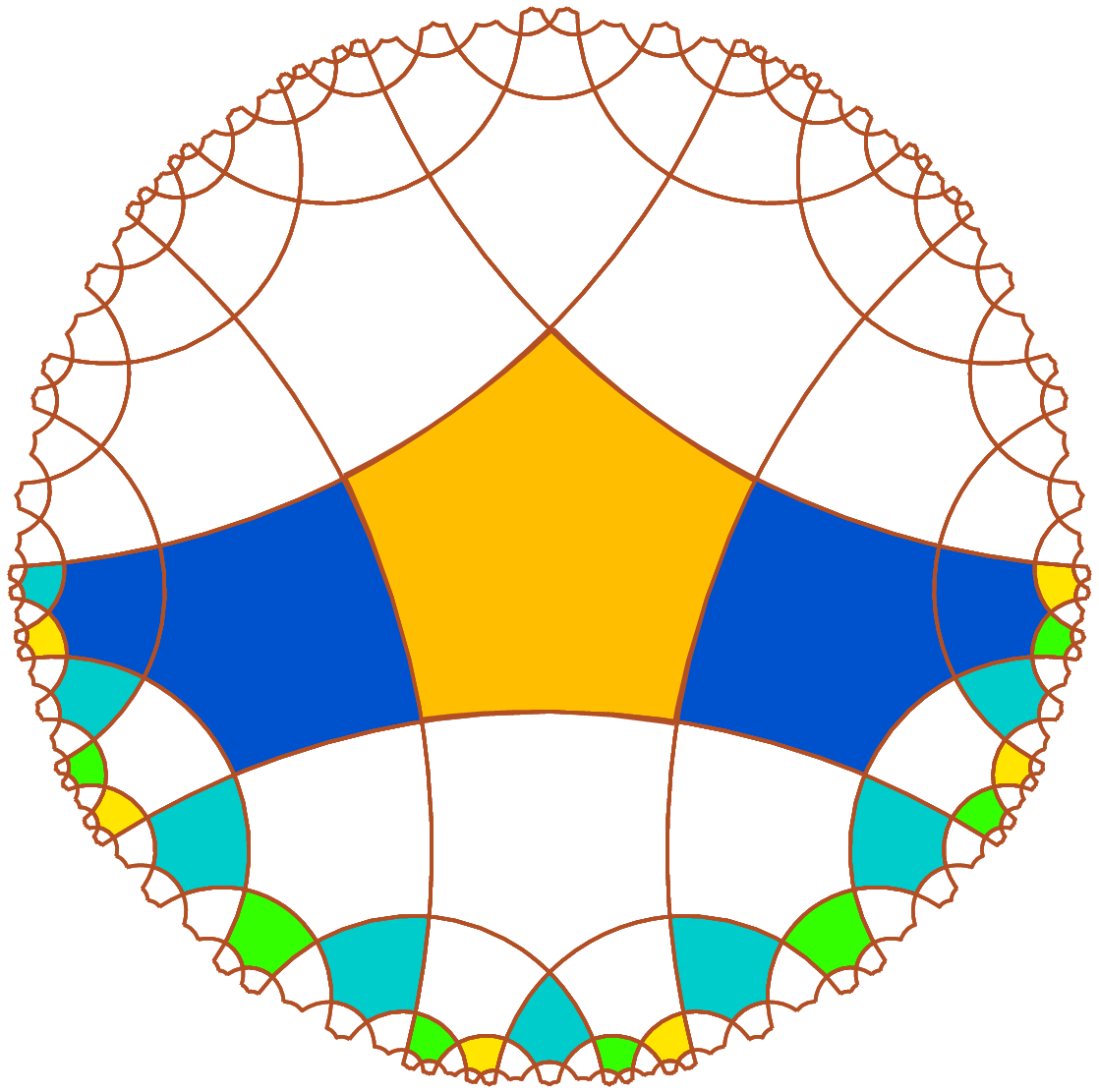,width=155pt}}
\setbox112=\hbox{\epsfig{file=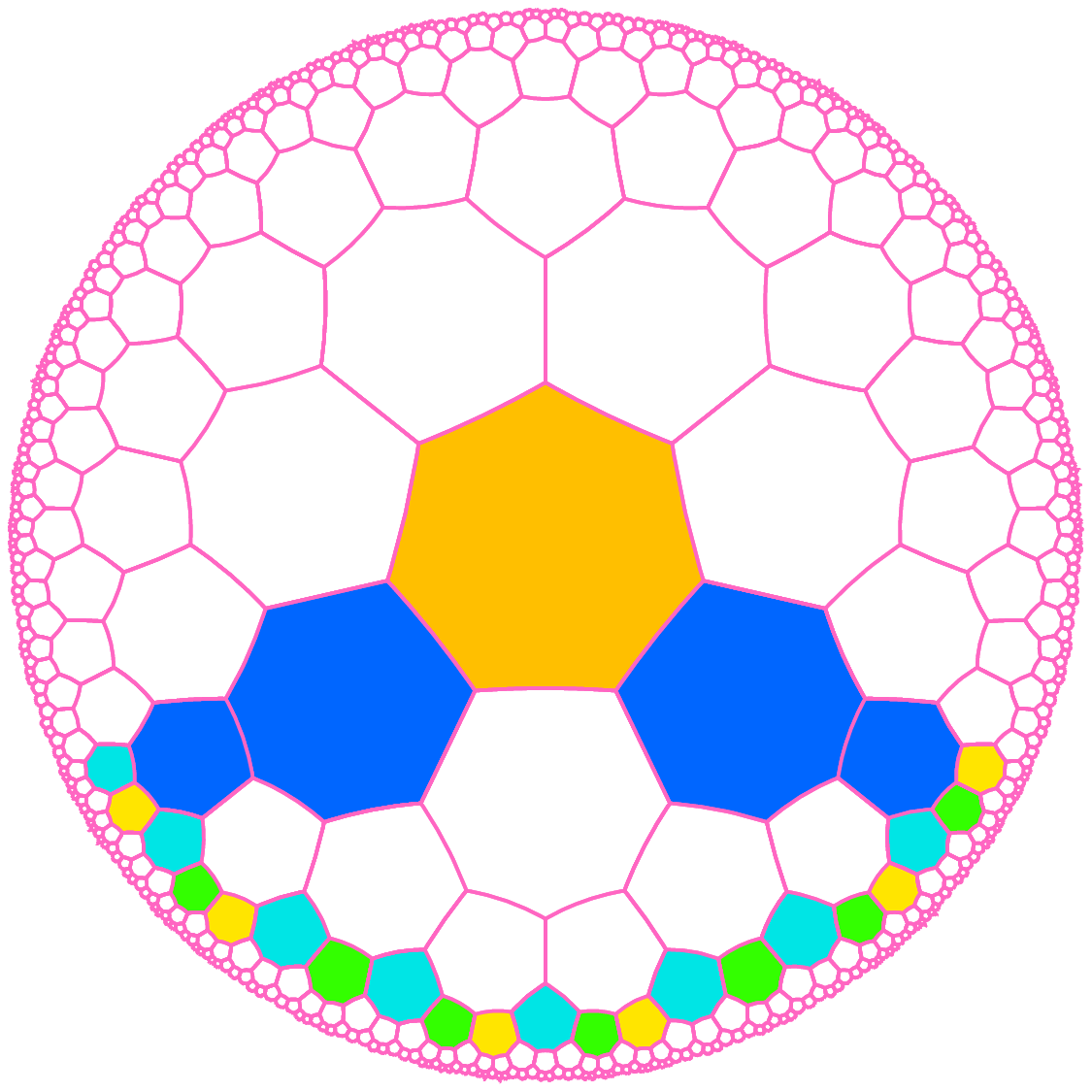,width=155pt}}
\vtop{
\ligne{\hfill
\PlacerEn {-155pt} {0pt} \box110
\PlacerEn {5pt} {0pt} \box112
\hfill
}
\begin{fig}\label{cont_Wsectors}
\leurre
The representation of $\partial W_3$.
Left-hand side: the pentagrid. Right-hand side: the heptagrid. In both cases,
the contour word is $rssbwbwwbwwss$.
\end{fig}
}

\ligne{\hfill}
\setbox110=\hbox{\epsfig{file=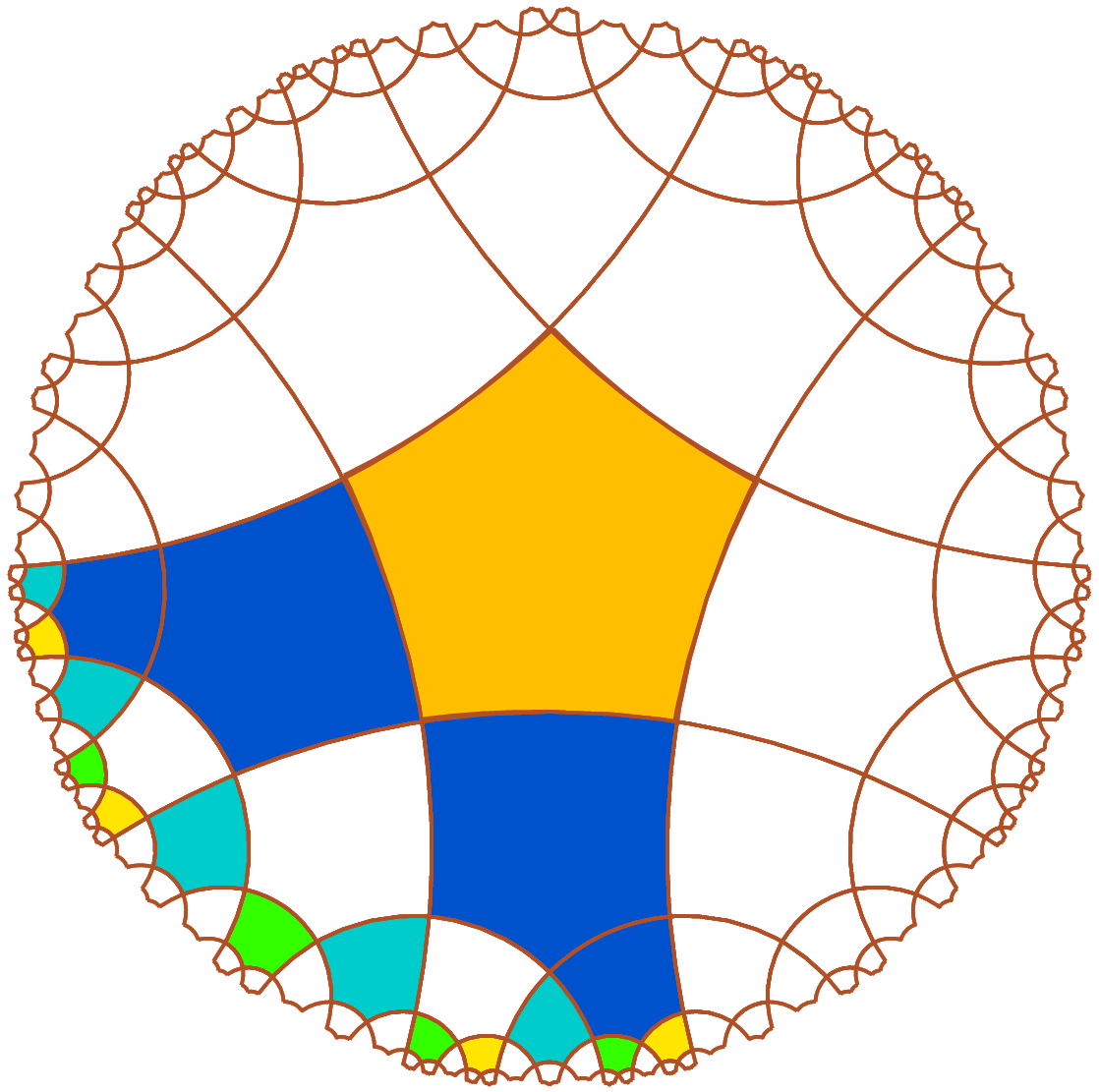,width=155pt}}
\setbox112=\hbox{\epsfig{file=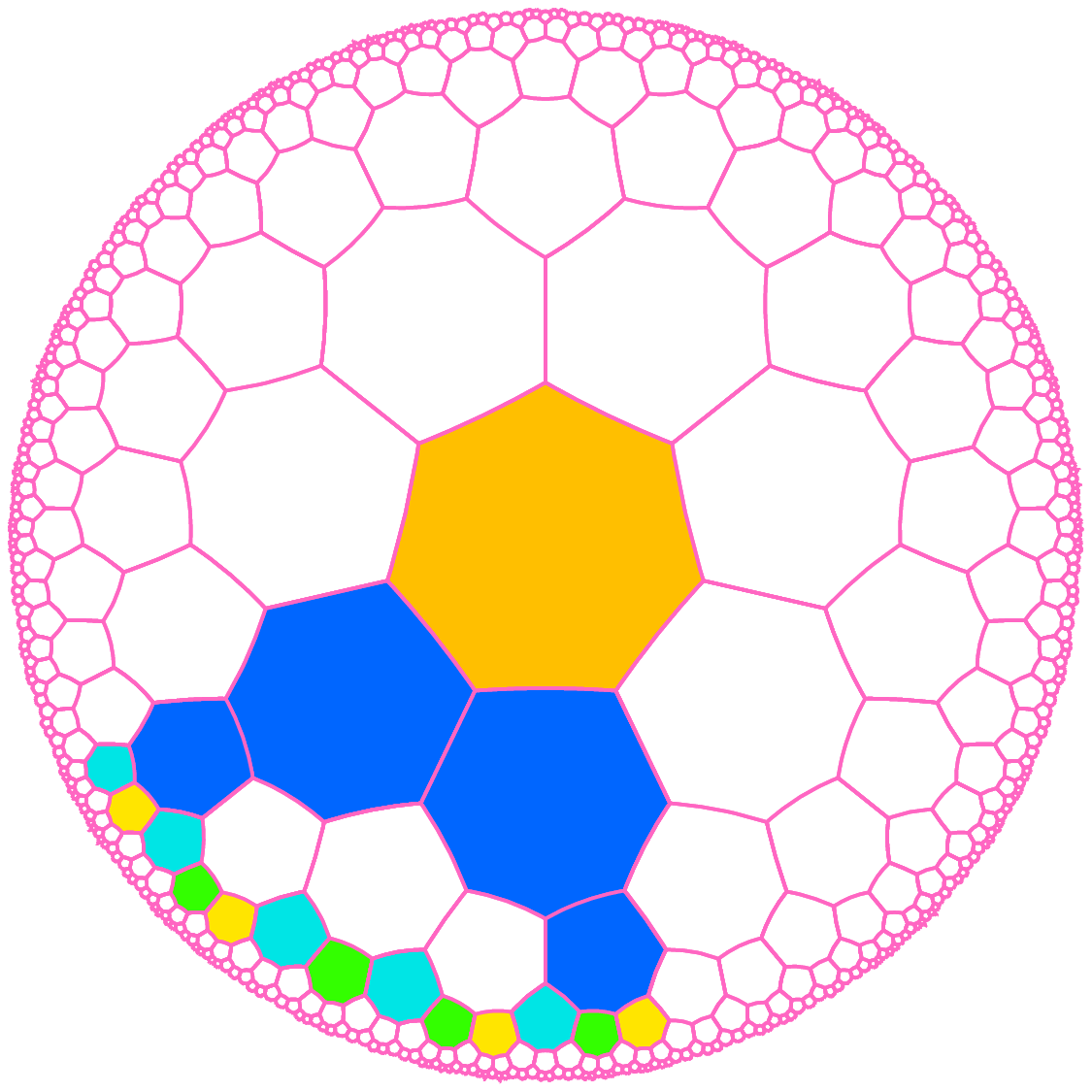,width=155pt}}
\vtop{
\ligne{\hfill
\PlacerEn {-155pt} {0pt} \box110
\PlacerEn {5pt} {0pt} \box112
\hfill
}
\begin{fig}\label{cont_Bsectors}
\leurre
The representation of $\partial B_3$.
Left-hand side: the pentagrid. Right-hand side: the heptagrid. In both cases,
the contour word is $rssbwbwwss$.
\end{fig}
}

\noindent 
Proof. There is no problem to recognize the leftmost branch of the tree, as it is always
on the top of the store. For the rightmost branch, the idea of the proof consists in 
introducing new rules which will in leave on the end of the external store a witness of 
each node on the rightmost branch. The rules can be defined as follows:

{\leftskip 20pt\parindent 0pt
$W_r \rightarrow B_bWW_bX$, $W_b\rightarrow BWW_bX$, $B_b \rightarrow B_bW$,\vskip 2pt
$W \rightarrow BWW$, $B \rightarrow BW$. 
\par}

Note that the second line contains the rules already used in automaton~\ref{autom2}.
The interpretation of the new rules is straightforward: $W_r$ stands for the root,
$B_b$ for the tiles on the leftmost branch below the root, which are black nodes,
$W_b$ for the tiles on the rightmost branch below the root. Now, $X$ is a witness left
by $W_b$ in order to remember the tile on the rightmost branch of the tree, look at
the corresponding instructions in automaton~\ref{autom3}.

   Note that automaton~\ref{autom3} recognizes the contour word of truncated white sectors.
For truncated black sectors, it is enough to replace the rule $W_r\rightarrow B_bWW_b$
by a rule $B_r\rightarrow B_bW_r$ and to replace the transitions

{\leftskip 20pt\parindent 0pt
$\delta(q_0,\epsilon,Z) = \{(q_0,push_2(F)),(q_0,push_1(W_r))\}$,\vskip 2pt
$\delta(q_0,\epsilon,ZF) = \{(q_0,push_2(FF)),(q_0,push_1(W_r))\}$
\par}

\noindent
by the transitions

{\leftskip 20pt\parindent 0pt
$\delta(q_0,\epsilon,Z) = \{(q_0,push_2(F)),(q_0,push_1(B_r))\}$,\vskip 2pt
$\delta(q_0,\epsilon,ZF) = \{(q_0,push_2(FF)),(q_0,push_1(B_r))\}$
\par}

\noindent
and to change accordingly the transitions involving $W_r$ by transitions involving~$B_r$.
   The proof of the correctness of automaton~\ref{autom3} and the modified automaton
for the truncated black sectors is straightforward and it is left to the reader. 
\vtop{
\begin{automaton}\label{autom3}
\leurre
The $2$-pushdown automaton recognizing the contour word of a truncated white sector in the 
pentagrid or in the heptagrid.
\end{automaton}
\vspace{-12pt}
\grostrait
\vskip 0pt
{\leftskip 20pt\parindent 0pt
two states: $q_0$ and $q_1$; input word in $\{r,s,b,w\}^*$; \vskip 0pt\noindent
$\Gamma = \{Z,B,W,B_b,W_b,W_r,X,F\}$;
\vskip 0pt\noindent
initial state: $q_0$; initial stack: $Z[\epsilon]$; transition function $\delta$: 
\par}
\vskip 0pt
\vspace{-4pt}
\demitrait
\vskip 0pt
{\leftskip 20pt\parindent 0pt
$\delta(q_0,\epsilon,Z) = \{(q_0,push_2(F)),(q_0,push_1(W_r))\}$\vskip 2pt
$\delta(q_0,\epsilon,ZF) = \{(q_0,push_2(FF)),(q_0,push_1(W_r))\}$\vskip 2pt
$\delta(q_0,\epsilon,WF) = (q_1,pop_2)$\vskip 2pt
$\delta(q_0,\epsilon,BF) = (q_1,pop_2)$\vskip 2pt
$\delta(q_0,r,W_rF) = (q_1,pop_2)$\vskip 2pt
$\delta(q_0,r,W_r) = (q_1,pop_1)$\vskip 2pt
$\delta(q_0,\epsilon,W_bF) = (q_1,pop_2)$\vskip 2pt
$\delta(q_0,s,B_bF) = (q_1,pop_2)$\vskip 2pt
$\delta(q_0,b,B_b) = (q_0,pop_1)$\vskip 2pt
$\delta(q_0,b,B) = (q_0,pop_1)$\vskip 2pt
$\delta(q_0,w,W) = (q_0,pop_1)$\vskip 2pt
$\delta(q_0,w,W_b) = (q_0,pop_1)$\vskip 2pt
$\delta(q_0,s,XF) = (q_0,pop_1)$\vskip 2pt
$\delta(q_1,\epsilon,W_rF) = (q_0,push_1(B_bWW_r))$\vskip 2pt
$\delta(q_1,\epsilon,W_bF) = (q_0,push_1(BWW_rX))$\vskip 2pt
$\delta(q_1,\epsilon,WF) = (q_0,push_1(BWW))$\vskip 2pt
$\delta(q_1,\epsilon,B_bF) = (q_0,push_1(B_bW))$\vskip 2pt
$\delta(q_1,\epsilon,BF) = (q_0,push_1(BW))$\vskip 2pt
\par}
\vskip 0pt
\demitrait
}
\vskip 10pt

\cqfd

\section{Dimensions 3 and 4}
  
   We take advantage of the remark we formulated after the proof of theorem~\ref{auto57}
in order to mention that the results of the previous section can be extended to the
hyperbolic $3D$ and~$4D$ spaces. As already mentioned in the introduction, there are
only a few tessellations in the hyperbolic~$3D$ and~$4D$ spaces and, starting from 
dimension~5, there is no such tiling in the hyperbolic space.

   We shall briefly indicate the reason why the result of theorem~\ref{auto57} can be extended
to two tilings of the hyperbolic $3D$ and~$4D$ spaces. We start with the $3D$~case first.

\subsection{Dimension $3$}

   There are four tessellations in the hyperbolic $3D$~space. We shall consider only
one of them, namely the tiling $\{5,3,4\}$, which we shall call the {\bf dodecagrid},
as it is built by tessellation starting from the regular dodecahedral with right angles.
The numbers in the signature $\{p,q,r\}$ say that the faces have $p$~sides, that,
on a tile, $q$~faces meet at a vertex and that any edge of a tile is shared by exactly
four tiles. Such a dodecahedron is unique up to isometries of the hyperbolic $3D$~space. 
In~\cite{mmgsFI,mmbook1}, it is proved that the dodecagrid can be split into eight 
sectors exactly, the eight sectors being defined by a {\it leading} dodecahedron.
The eight leading dodecahedron share a common vertex called the {\bf central point} and 
each one is in contact with exactly three of them through three of its faces. These 
faces of contact define three planes of the hyperbolic $3D$~space which have a common 
point and which are pairwise perpendiculat. Now, each sector is spanned by a tree~$\cal T$ 
whose root is associated to its leading dodecahedron. 

   Now, the tree $\cal T$ is a $3D$-one which can also be represented as a planar one 
as its generation can be given by finitely many rules looking very much to those used in the
case of the pentagrid or the heptagrid. Without entering in further details about the 
justification of this property, see~\cite{mmgsFI,mmbook1} for such a study, we can 
indicate a set of four rules which generate~$\cal T$ on the basis of the labelling
of each node with one of the letters $O$, $D$, $C$ and~$T$, the leading dodecahedron beling
labelled with~$O$:

\def\ttV{\vrule depth 6pt height 12pt width 0pt}
\def\ttH{\hrule depth 0pt height 0pt width \hsize}

\def\laligne #1 #2 #3 #4 #5 {%
\ligne{\ttV\hbox to 35pt{#1\hfill}
       \ttV\hbox to 20pt{\hfill #2\hfill}
       \ttV\hbox to 20pt{\hfill #3\hfill}
       \ttV\hbox to 20pt{\hfill #4\hfill}
       \ttV\hbox to 20pt{\hfill #5\hfill}
       \ttV}
\ttH
}
\vskip 5pt
\ligne{\hfill
$O \rightarrow O^5C^3T$, $H \rightarrow O^4C^3T$, $C \rightarrow O^3C^3T$,
$T \rightarrow O^2HC^2T$
\hfill$(R_3)$\hskip 20pt}
\vskip 5pt
Note that the rules can be given a matricial representation as indicated in 
Table~\ref{matrix3} with self-explaining notations.

   From this, using the already applied technique of the previous paragraph, it is possible
to devise a $2$-iterated pushdown automaton which recognize the contour words of a region
whose definition is a bit changed with respect to what it was given in the previous
section. Here, a ball~$B^3_k$ of radius~$k$ is the set of tiles which are within a 
distance~$k$, in tiles, from the central point. Define the border of~$B^3_k$, 
denoted by~$\partial B^3_k$, as the tiles which are at a distancce~$k$
exactly from the central point. The contour word is defined by traversing $\partial B^3_k$
in the way induced by the maps defined in~\cite{mmgsFI,mmbook1}. Indeed, there is a way
to injectively enumerate the tiles of the different level of~$\cal T$ which allow
to lift up a planar representation of~$\cal T$ as defined by the rules~$(R_3)$ up to
the actual $3D$-tree. This can be obtained by maps of level~1 and there is
a way, described in\cite{mmgsFI,mmbook1} to generate the map of the level~$k$ from the maps
of level~1 of the different trees defined by the rules~$(R_3)$, considering the trees
rooted at each possible kind of nodes.

\vtop{
\begin{tab}\label{matrix3}
\leurre
                             The matrix of the generating rules of dodecagrid:
\end{tab}
\vspace{-14pt}
\grostrait
\vspace{-8pt}
\vskip 0pt
\ligne{\hfill\vtop{\offinterlineskip\leftskip 0pt\parindent0pt\hsize=125pt
\ttH  \laligne {} {$O$} {$H$} {$C$} {$T$}
}
\hfill}
\demitrait
\ligne{\hfill\vtop{\offinterlineskip\leftskip 0pt\parindent0pt\hsize=125pt
\ttH  \laligne {$O$} 5 0 3 1
\ttH  \laligne {$H$} 4 0 3 1
\ttH  \laligne {$C$} 3 0 3 1
\ttH  \laligne {$T$} 2 1 2 1
\ttH
}
\hfill}
\vskip 3pt
\demitrait
}
\vskip 10pt
   The sectors themselves do generate a contour word. The just indicated maps can be used
to identify the {\bf sides} of the tree in the maps and the considered nodes of the
tree can be given an appropriate sign. Then, the traversal defined by the extension
of automaton~\ref{autom2} to this case will generate a contour word in which we
have first the signs corresponding to the nodes which are on the sides of the sector
and then the nodes which are at distance~$k$, using the traversal defined by the maps.
As we have four possible labels for the nodes, the rules~$(R_3)$ define four kinds
of sectors. We shall call {\bf $k$-truncated $\gamma$-sector}, with $\gamma\in\{O,H,C,T\}$, 
the set of tiles which belong to a tree generated by the rules~$(R_3)$ which is rooted 
at a tile labelled with~$\gamma$.

   Then we have:

\begin{thm}\label{3D}
There is a $2$-iterated pushown automaton which recognizes the contour words of the 
balls~$B_k$ in the hyperbolic $3D$-space.
Also, for each $\gamma\in\{O,H,C,T\}$, there is a $2$-iterated psushdown automaton which
recognizes the contour words of any $k$-truncated $\gamma$-sector.
\end{thm}

\subsection{Dimension $4$}

   In the hyperbolic~$4D$ space, there are five tessellations based on a
regular polytope. We shall take the one which extends the dodecagrid. From the regular
dodecahedron with right angles, it is possible to construct a regular polytope called
the 120-cell, whose faces are regular dodecahedra with right angles. From this regular
polytope, we can generate a tiling by tessellation which we call the {\bf 120-cell grid}.
The signature of this grid is $\{5,3,3,4\}$ which means that four 120-cells share a common
pentagon, that three dodecahedra share a common edge, and that three pentagons share a 
common point, five remembring that the 2-dimensional structure is a pentagon.
   
   The common point with the dodecagrid is that the technique used for the dodecagrid also
applies here. The space is the union of 16~sectors, each one having a leading 120-cell
sharing a common vertex which is the central point. Each sector is spanned by a tree
whose root is associated to the leading 120-cell. This tree is also generated by a finite
set of rules of the same type as that of the rules~$(R_3)$. Now, this time the set of 
rules is much more complex as it involves 11 labels. For this reason, we shall use the 
matricial representation of the previous subsection with a slight modification. In 
Table~\ref{matrix4}, the leftmost column indicate the labels.
The labels are not repeated on the first line: it is assumed that the coefficients of the
matrix apply to the type of node whose label is on the row whose index is the same as the
column index of the coefficient, the indices starting from~1 and the leftmost column
receiving index~0. We refer the reader to~\cite{mmJUCS4D} for the justification of these
rules and for further explanations.

\newdimen\largeur\largeur=24pt
\def\premdemiligne #1 #2 #3 #4 #5 #6
          {\ttV\hbox to \largeur{\bf\hfill#1\hfill}\ttV
           \ttV\hbox to \largeur{\hfill#2\hfill}
           \ttV\hbox to \largeur{\hfill#3\hfill}
           \ttV\hbox to \largeur{\hfill#4\hfill}
           \ttV\hbox to \largeur{\hfill#5\hfill}
           \ttV\hbox to \largeur{\hfill#6\hfill}
          }
\def\deuzedemiligne #1 #2 #3 #4 #5 #6
          {\ttV\hbox to \largeur{\hfill#1\hfill}
           \ttV\hbox to \largeur{\hfill#2\hfill}
           \ttV\hbox to \largeur{\hfill#3\hfill}
           \ttV\hbox to \largeur{\hfill#4\hfill}
           \ttV\hbox to \largeur{\hfill#5\hfill}
           \ttV\hbox to \largeur{\hfill#6\hfill}
          }

\vtop{
\begin{tab}\label{matrix4}
\leurre
                             The matrix of the generating rules of the 120-cell grid:
\end{tab}
\vspace{-14pt}
\grostrait
\vspace{-8pt}
\ligne{\hfill \vtop{\offinterlineskip\parindent0pt\leftskip0pt
                    \hsize=323pt
    \ttH
    \ligne{%
    \premdemiligne {9} 6 {10} {21} {35} 3
    \deuzedemiligne {19} {14} 5 1 1 1
    \ttV}\ttH
    \ligne{%
    \premdemiligne {8} 5 {10} {21} {35} 3
    \deuzedemiligne {19} {14} 5 1 1 1
    \ttV}\ttH
    \ligne{%
    \premdemiligne {7} 4 {10} {21} {35} 3
    \deuzedemiligne {19} {14} 5 1 1 1
    \ttV}\ttH
    \ligne{%
    \premdemiligne {6$a$} 3 {11} {20} {35} 3
    \deuzedemiligne {19} {14} 5 1 1 1
    \ttV}\ttH
    \ligne{%
    \premdemiligne {6$b$} 2 {12} {20} {35} 3
    \deuzedemiligne {19} {14} 5 1 1 1
    \ttV}\ttH
    \ligne{%
    \premdemiligne {5} 2 {11} {20} {35} 3
    \deuzedemiligne {19} {14} 5 1 1 1
    \ttV}\ttH
    \ligne{%
    \premdemiligne {4} 2 {10} {20} {35} 3
    \deuzedemiligne {19} {14} 5 1 1 1
    \ttV}\ttH
    \ligne{%
    \premdemiligne {3} 1 {11} {19} {35} 3
    \deuzedemiligne {19} {14} 5 1 1 1
    \ttV}\ttH
    \ligne{%
    \premdemiligne {2} 1 {10} {19} {35} 3
    \deuzedemiligne {19} {14} 5 1 1 1
    \ttV}\ttH
    \ligne{%
    \premdemiligne {1} 1 {10} {18} {35} 3
    \deuzedemiligne {19} {14} 5 1 1 1
    \ttV}\ttH
    \ligne{%
    \premdemiligne {0} 1 {10} {18} {34} 3
    \deuzedemiligne {19} {14} 5 1 1 1
    \ttV}\ttH
   }
       \hfill}
\vspace{1pt}
\demitrait
}
\vskip 10pt
   Defining the balls by the distance from a central point as in the $3D$~case, and the
sectors
   From table~\ref{matrix4}, we can devise the instructions of a 2-iterated pushdown automaton
which will recognize the contour word of a ball. We also can do the same for the
$k$-truncated $\gamma$-sectors with 
$\gamma\in\{${\bf 9$,$8$,$7$,$6$a,$6$b,$5$,$4$,$3$,$2$,$1$,$0}$\}$

\section*{Acknowledgment}

I wish to express special thanks to G\'eraud S\'enizergues for drawing my attention on
iterated pushdown automata and for his interest in the results of this paper.


\begin{thebibliography}{5}
\def\kvs{\vspace{-1pt}}
\parskip -1pt

\bibitem{fratani-senizergues}
Fratani~S., S\'enizergues~G.,
Iterated pushdown automata and sequences of rational numbers,
{\it Annals of pure and applied logic}, {\bf 141}, (2006), 363-411. 
\kvs
\bibitem{greibach}
S. Greibach,
Full AFL's and nested iterated substitution,
{\it Information and Control}, {\bf 16}(1), (1970), 7-35.
\kvs

\bibitem{mmJUCSii}
M. Margenstern,
New Tools for Cellular Automata of the Hyperbolic Plane,
{\it Journal of Universal Computer Science}
{\bf 6}(12), (2000), 1226--1252.
\kvs
\bibitem{mmJUCS4D}
M. Margenstern,
The tiling of the hyperbolic $4D$ space by the 120-cell is
combinatoric,
{\it Journal of Universal Computer Science},
 {\bf 10}(9), (2004), 1212-1238.
\kvs
\bibitem{mmbook1}
Margenstern~M.,
Cellular Automata in Hyperbolic Spaces, Volume 1, Theory,
{\it OCP}, Philadelphia, (2007), 422p.

\bibitem{mmbook2}
Margenstern~M.,
Cellular Automata in Hyperbolic Spaces, Volume 2, Implementation and computations,
{\it OCP}, Philadelphia, (2003), 360p.
\kvs
%
%
%
\bibitem{mmgsJUCS}
M. Margenstern, G. Skordev,
Fibonacci Type Coding for the Regular Rectangular Tilings of the
Hyperbolic Plane,
{\it Journal of Universal Computer Science},
{\bf 9}(5), (2003), 398-422.

\bibitem{mmgsFI}
M. Margenstern, G. Skordev,
Tools for devising cellular automata in the hyperbolic 3D space,
{\it Fundamenta Informaticae}, {\bf 58}(2), (2003), 369-398.

\kvs
\bibitem{maslov}
Maslov A. N.,
The hierarchy of indexed languages, {\it Soviet Mathematics, Doklady}, {\bf 15}, (1974),
1170-1174.
%

\end{thebibliography}
\end{document}